\documentclass[conference]{IEEEtran}
\IEEEoverridecommandlockouts
\usepackage{cite}
\usepackage{amsmath,amssymb,amsfonts}
\usepackage{algorithmic}
\usepackage{graphicx}
\usepackage{textcomp}
\usepackage{xcolor}
\def\BibTeX{{\rm B\kern-.05em{\sc i\kern-.025em b}\kern-.08em
    T\kern-.1667em\lower.7ex\hbox{E}\kern-.125emX}}
\usepackage{booktabs}
\newcounter{algorithm}
\begin{document}
\raggedbottom

\title{FAPO: \textbf{F}anout-\textbf{A}ware \textbf{P}ost-Mapping \textbf{O}ptimization for LUT-Based FPGAs}

\author{\IEEEauthorblockN{Xiaoyu Hao, Keren Zhu\textsuperscript{*}}
\IEEEauthorblockA{\textit{State Key Lab of Integrated Chips and Systems, Fudan University, China}\\
xyhao26@m.fudan.edu.cn, krzhu@fudan.edu.cn}
\thanks{\textsuperscript{*}Corresponding author: Keren Zhu.}}

\maketitle

\begin{abstract}
Reducing logic depth during FPGA technology mapping can leave fanout bottlenecks in the resulting LUT network.
LUT replication redistributes these loads but consumes area and increases loading on shared fanins.
We propose FAPO, a mapper-independent post-mapping optimizer that jointly refines sink assignments and LUT cuts under a final-area constraint.
It evaluates replication, cut switching, and recovery by the resulting live LUT count and full-network timing.
A necessary all-critical-path condition motivates minimum-cut proposals that complement greedy refinement.
Within a fixed critical graph and a restricted same-cut action family, each proposal minimizes the number of copy actions needed to cover all current critical paths.
Experiments on EPFL benchmarks show that FAPO reduces the geometric-mean fanout-aware area--timing product (ATP) by up to 14.48\% with only a small increase in area.
VPR evaluation gives a 6.82\% geometric-mean routed-delay reduction relative to ABC \texttt{if}.
\end{abstract}

\begin{IEEEkeywords}
FPGA technology mapping, lookup tables, post-mapping optimization, fanout-aware timing, area--timing product
\end{IEEEkeywords}

\section{Introduction}
\emph{LUT technology mapping} implements a logic network with lookup tables (LUTs) and determines the logic area and depth of a field-programmable gate array (FPGA) design. Classical methods optimize depth and area through network-flow and cut-based techniques \cite{CongDing1994FlowMap,CongDing1994AreaDepth,MishchenkoEtAl2007PriorityCuts,CongWuDing1999CutRanking,MishchenkoEtAl2007Mapping,ChenCong2004DAOmap}. Depth alone, however, does not capture fanout-dependent delay \cite{CongDing1994NominalDelay}. Networks with equal logic depth can impose different loads on their LUTs. A high-fanout LUT can therefore remain a timing bottleneck after depth optimization. Placement-driven mapping uses physical delay estimates \cite{LinEtAl2003PDM}, while WireMap reduces pin-to-pin connections to improve routability \cite{JangEtAl2009WireMap}.

Recent methods expand the mapping search space through adaptive gate decomposition, Ashenhurst--Curtis decomposition, equality saturation, and heterogeneous representations \cite{FanWu2023AGDMap,TempiaCalvinoEtAl2025ACD,HofmannEtAl2025EqMap,HuEtAl2025MCH}. Resubstitution and mapping fusion further improve mapping quality \cite{CostamagnaEtAl2025Resub,Yu2025FuseMap}. Their resulting LUT covers can still contain concentrated sink loads. A legal cover exposes sink assignments and selected cuts as a common interface for further refinement, independent of the baseline mapper's internal representation.

Reducing fanout under an area constraint requires coordinating replication with the surrounding mapping. Prior work addresses fanout through gate duplication \cite{ChenTsui1999Duplication} and placement-coupled FPGA replication \cite{BeraudoLillis2003Replication}. Replication splits sink loads but increases loading on shared fanins. Cut switching changes these dependencies, while cleanup and recovery remove redundant instances. FAPO evaluates these changes jointly and measures area by the final live LUT count, allowing recovered area to offset replication costs.

Coordinating replication also matters when several paths share the maximum delay. Improving one path can leave another unchanged. FAPO combines greedy joint refinement with minimum-cut proposals that cover every current critical path. Full-network reevaluation accounts for upstream loading and newly exposed bottlenecks. Our main contributions are as follows:
\begin{itemize}
\item We formulate joint sink-assignment and LUT-cut optimization on an imported LUT cover, with replication, cut switching, and recovery constrained by final live LUT count.
\item We derive a necessary all-critical-path condition and use minimum cut to find a minimum-cardinality covering copy set on a fixed critical graph within the same-cut action family.
\item We augment greedy refinement with at most one minimum-cut proposal per sweep and select candidates using recomputed network area and timing.
\item On EPFL benchmarks, FAPO reduces geometric-mean ATP by up to 14.48\% with a 5\% area allowance. VPR results show a 6.82\% geometric-mean reduction in routed delay.
\end{itemize}

\begin{figure*}[t]
\centering
\includegraphics[width=\textwidth]{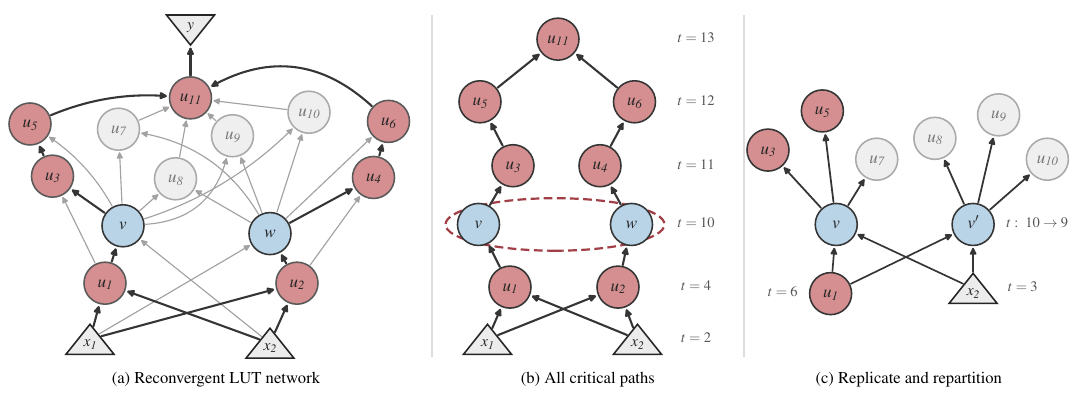}
\caption{Illustrative refinement ($K=6$, $\alpha=1$). Triangles denote inputs $x_i$ and output $y$; circles denote LUT instances. (a) Reconvergent LUT network; critical arcs are dark. (b) Arrivals computed on (a) identify four longest paths. The red dashed contour marks the covering set $\{v,w\}$. (c) Local sink repartition between $v$ and its copy $v'$; other fanins are omitted. The same operation applies to $w$, followed by full-network area and timing evaluation.}
\label{fig:flow}
\end{figure*}

\begin{figure}[t]
\centering
\begin{minipage}{\columnwidth}
\hrule\smallskip
\refstepcounter{algorithm}\label{alg:fapo}
\textbf{Algorithm \thealgorithm: FAPO post-mapping refinement}\par\smallskip
\footnotesize
\begin{algorithmic}[1]
\REQUIRE Legal seed $M_0$, $\alpha$, $\beta$
\ENSURE Area-feasible archive $\mathcal P$ and selected mapping $M^*$
\STATE Set $A_{\max}$ by \eqref{eq:area}; initialize $\mathcal F$ and $\mathcal A$ from $M_0$
\REPEAT
\STATE $\mathcal Q\gets\operatorname{Evaluate}(\operatorname{Expand}(\mathcal F),A_{\max})$
\STATE $\mathcal R\gets\varnothing$
\IF{the cut branch is enabled}
\STATE $\widehat M\gets\operatorname{ATPBest}(\operatorname{ParetoSample}(\mathcal Q))$
\STATE $M_c\gets\operatorname{CutBundle}(\widehat M,A_{\max})$
\STATE $\mathcal R\gets\operatorname{AdmitCut}(M_c,\widehat M,A_{\max})$
\ENDIF
\STATE $\mathcal A\gets\operatorname{UpdateArchive}(\mathcal A\cup\mathcal Q\cup\mathcal R)$
\STATE $\mathcal F\gets\operatorname{UpdateFrontier}(\mathcal F,\mathcal Q)$
\UNTIL{convergence or the search limit is reached}
\STATE $\mathcal P\gets\operatorname{Finalize}(\mathcal A)$
\RETURN $M^*=\arg\min_{M\in\{M_0\}\cup\mathcal P} A(M)T_\alpha(M)$, $\mathcal P$
\end{algorithmic}
\smallskip\hrule
\end{minipage}
\end{figure}

\section{Mapped-Network Optimization}
\subsection{State and final-area constraint}
Let $M_0=\mathcal{B}(G)$ be a legal $K$-LUT mapping of a combinational source network $G$. The optimizer imports $M_0$, while $G$ specifies the reference function for equivalence checking \cite{MishchenkoEtAl2006CEC}. A \emph{live instance} is a LUT reachable backward from a primary output. We represent it by $(r,C,S)$, where $r$ is its logic root, $C$ its selected $K$-feasible cut, and $S$ its served sinks. A \emph{$K$-feasible cut} has at most $K$ leaves that determine the root function. A \emph{sink} is a LUT input or primary-output connection. Copies of a root serve disjoint sink sets and may use different cuts. Cleanup removes unreachable instances.

Area $A(M)$ counts live LUT instances after cleanup and recovery. Given $A_0=A(M_0)$ and allowance $\beta$, the output must satisfy
\begin{equation}
A(M)\leq A_{\max}=A_0+\lfloor\beta A_0\rfloor.
\label{eq:area}
\end{equation}
The bound applies to the final live network, so cleanup and recovery can offset replication costs.

FAPO builds its refinement graph directly from the imported LUT cover, without the baseline mapper's internal subject graph or cut database.

\subsection{Fanout-aware timing and objective}
The \emph{fanout} $\operatorname{fo}_M(u)$ counts live LUT-input arcs and primary-output sinks driven by $u$. For penalty $g(f)=\max(f-1,0)$ and weight $\alpha\geq0$, arrivals at primary inputs $p$ and LUTs $I$ are
\begin{align}
t(p)&=\alpha g(\operatorname{fo}_M(p)),\label{eq:pi}\\
t(I)&=1+\alpha g(\operatorname{fo}_M(I))
       +\max_{J\in\operatorname{fanin}(I)}t(J).\label{eq:arrival}
\end{align}
The maximum output-driver arrival is $T_\alpha(M)$. This mapping-level proxy combines logic depth with a fanout penalty. The search targets
\begin{equation}
\min_{M\equiv G,\,A(M)\leq A_{\max}} A(M)T_\alpha(M),
\label{eq:objective}
\end{equation}
the \emph{fanout-aware area--timing product} (ATP). The search retains nondominated $(A,T_\alpha)$ pairs to explore area--delay tradeoffs. The original seed remains a protected candidate. Feasible mappings preserve acyclicity, $K$-feasibility, and functional equivalence within the area cap.

\begin{table*}[t]
\caption{EPFL mapping results at $K=6$, $\alpha=1$, and $\beta=5\%$. Initial and FAPO columns report $A/T_1$ before and after optimization. ATP ratios are FAPO/initial; bold values indicate improvements.}
\label{tab:parents}
\centering\footnotesize
\setlength{\tabcolsep}{3pt}
\begin{tabular}{lrrrrrrrrr}
\toprule
& \multicolumn{3}{c}{ABC \texttt{if}} & \multicolumn{3}{c}{ACD-remap} & \multicolumn{3}{c}{AGDMap-D}\\
\cmidrule(lr){2-4}\cmidrule(lr){5-7}\cmidrule(lr){8-10}
Circuit & Initial & FAPO & ATP ratio & Initial & FAPO & ATP ratio & Initial & FAPO & ATP ratio\\
\midrule
adder & 254/163 & 266/135 & \textbf{0.8673} & 327/68 & 338/43 & \textbf{0.6536} & 254/179 & 266/150 & \textbf{0.8776} \\
bar & 512/138 & 512/138 & 1.0000 & 512/138 & 512/138 & 1.0000 & 561/274 & 564/236 & \textbf{0.8659} \\
div & 22113/6483 & 22132/6052 & \textbf{0.9343} & 7044/3819 & 7073/3151 & \textbf{0.8285} & 19156/9210 & 19174/8746 & \textbf{0.9505} \\
log2 & 8092/1091 & 8118/905 & \textbf{0.8322} & 8661/1071 & 8690/921 & \textbf{0.8628} & 7472/1210 & 7496/1035 & \textbf{0.8581} \\
max & 769/304 & 774/222 & \textbf{0.7350} & 786/381 & 793/157 & \textbf{0.4157} & 798/367 & 809/248 & \textbf{0.6851} \\
multiplier & 5927/424 & 5954/403 & \textbf{0.9548} & 6386/400 & 6406/373 & \textbf{0.9354} & 5376/598 & 5367/446 & \textbf{0.7446} \\
sin & 1464/549 & 1488/426 & \textbf{0.7887} & 1812/521 & 1827/387 & \textbf{0.7490} & 1395/710 & 1418/534 & \textbf{0.7645} \\
sqrt & 5711/11796 & 5740/9577 & \textbf{0.8160} & 5963/9526 & 5992/7398 & \textbf{0.7804} & 6424/11482 & 6450/8903 & \textbf{0.7785} \\
square & 3998/299 & 4030/273 & \textbf{0.9204} & 4135/268 & 4150/246 & \textbf{0.9212} & 3592/359 & 3630/308 & \textbf{0.8670} \\
arbiter & 2722/68 & 2722/68 & 1.0000 & 2227/45 & 2233/43 & \textbf{0.9581} & 2715/68 & 2715/68 & 1.0000 \\
cavlc & 116/71 & 116/71 & 1.0000 & 117/68 & 117/67 & \textbf{0.9853} & 113/71 & 113/71 & 1.0000 \\
ctrl & 29/28 & 29/27 & \textbf{0.9643} & 29/27 & 29/27 & 1.0000 & 29/29 & 29/29 & 1.0000 \\
dec & 287/47 & 287/47 & 1.0000 & 284/92 & 284/92 & 1.0000 & 272/248 & 272/248 & 1.0000 \\
i2c & 353/116 & 356/93 & \textbf{0.8085} & 296/120 & 299/95 & \textbf{0.7997} & 348/116 & 351/107 & \textbf{0.9304} \\
int2float & 51/31 & 51/31 & 1.0000 & 45/28 & 46/27 & \textbf{0.9857} & 45/27 & 45/27 & 1.0000 \\
mem\_ctrl & 12084/1192 & 12102/1074 & \textbf{0.9023} & 10236/907 & 10255/778 & \textbf{0.8594} & 12104/1730 & 12124/1370 & \textbf{0.7932} \\
priority & 210/66 & 211/65 & \textbf{0.9895} & 177/19 & 177/18 & \textbf{0.9474} & 227/71 & 220/66 & \textbf{0.9009} \\
router & 89/31 & 83/17 & \textbf{0.5114} & 76/16 & 78/14 & \textbf{0.8980} & 74/20 & 77/17 & \textbf{0.8845} \\
voter & 2695/72 & 2709/63 & \textbf{0.8795} & 1728/45 & 1738/43 & \textbf{0.9611} & 2061/66 & 2066/60 & \textbf{0.9113} \\
\midrule
Geo. mean ATP ratio & & & 0.8798 & & & 0.8552 & & & 0.8796\\
\bottomrule
\end{tabular}
\end{table*}

\section{FAPO Algorithm}
\subsection{Coupled refinement under a final-area bound}
FAPO coordinates sink assignments and cuts because replication changes both output loads and upstream dependencies. It prioritizes LUTs by timing criticality and then fanout. For each selected LUT, it considers a bounded set of sink bipartitions, including balanced splits and single-sink moves. Each partition is paired with alternative $K$-feasible cuts enumerated on the imported-cover refinement graph. The original and copied instances may use different cuts, allowing load redistribution and upstream dependencies to change together.

Cleanup removes instances made unreachable by these changes. \emph{Copy recovery} merges same-root instances by redirecting their sinks to a surviving realization, retaining only legal merges. Because replication increases fanin loads and merging increases survivor loads, FAPO remeasures the resulting network using \eqref{eq:pi}--\eqref{eq:arrival}. The resulting live LUT count and delay determine feasibility under \eqref{eq:area} and admission to the search. Nondominated candidates remain available for exploration. Final selection minimizes ATP after recovery.

\subsection{Coordinated replication across critical paths}
For fixed $M$, the timing recurrence defines a node-weighted longest-path problem. Input weights are $\alpha g(\operatorname{fo}_M)$; LUT weights additionally include one logic-delay unit. Writing these weights as $w_M(u)$ gives
\begin{equation}
T_\alpha(M)=\max_{P}\sum_{u\in P}w_M(u),
\label{eq:path}
\end{equation}
where $P$ ranges over input-to-output paths. Forward arrival times and reverse remaining-path delays identify the \emph{critical directed acyclic graph} (DAG) $D^*(M)$. It contains all maximum-weight paths.

\textbf{Coverage condition.} Any path whose connections and node weights remain unchanged retains its delay. Strictly reducing $T_\alpha$ therefore requires affecting every current critical path. This necessary condition motivates coordinated copy sets across $D^*(M)$.

A \emph{same-cut action} partitions an eligible LUT's sinks into two nonempty sets served by the original and one copy with identical root, cut, and fanins. Each action introduces one LUT before recovery.

\textbf{Minimum-copy coverage.} FAPO selects coordinated copy actions through a minimum vertex cut on $D^*(M)$. Each vertex is split into an input and an output vertex, with unit capacity on eligible vertex arcs and infinite capacity elsewhere. A supersource and supersink connect the critical sources and outputs. Any finite cut intersects every current critical path, and its capacity counts the selected copy actions. When a finite cut exists, minimizing its capacity gives a smallest covering copy set within this fixed-DAG, same-cut family. A shared vertex can cover several reconvergent critical paths with one copy. The cut identifies replication sites. Sink repartition at these sites redistributes the loads between the original instances and their copies.

In Fig.~\ref{fig:flow}(b), $\{v,w\}$ covers all four longest paths; Fig.~\ref{fig:flow}(c) illustrates sink repartition at a selected LUT. Added fanin loads can expose new bottlenecks, so coverage alone does not ensure delay reduction. $\operatorname{CutBundle}$ materializes the selected copies, remeasures timing, and rebuilds the critical DAG after accepted sets. Its bounded rounds preserve feasibility and nonincreasing delay, producing a candidate for joint area--timing evaluation.

\begin{table}[t]
\caption{Search-strategy comparison at $\beta=5\%$. Ratios are geometric means relative to ACD-remap seeds at minimum ATP.}
\label{tab:ablation}
\centering\footnotesize
\setlength{\tabcolsep}{3pt}
\begin{tabular}{lrrr}
\toprule
Search strategy & ATP ratio & $T_1$ ratio & $A$ ratio\\
\midrule
Greedy-only (matched) & 0.8607 & 0.8549 & 1.0067\\
Minimum-cut-only & 0.8757 & 0.8704 & 1.0061\\
Minimum-cut + region refinement & 0.8753 & 0.8700 & 1.0061\\
50/50 greedy--region portfolio & 0.8682 & 0.8627 & 1.0064\\
\textbf{FAPO: greedy + one cut proposal} & \textbf{0.8553} & \textbf{0.8490} & \textbf{1.0074}\\
\bottomrule
\end{tabular}
\end{table}

\begin{table*}[t]
\caption{Routed critical-path delay (CPD) on 14 circuits with strict mapping-level ATP improvements under full FAPO ($\beta=5\%$). Red. denotes reduction. Positive reductions are bold.}
\label{tab:physical}
\centering\footnotesize
\setlength{\tabcolsep}{5pt}
\begin{tabular}{lrrr@{\hspace{22pt}}lrrr}
\toprule
Circuit & Initial (ns) & Refined (ns) & Red. (\%) & Circuit & Initial (ns) & Refined (ns) & Red. (\%)\\
\midrule
\texttt{adder} & 23.4439 & 22.5086 & \textbf{3.99} & \texttt{multiplier} & 33.1260 & 31.5619 & \textbf{4.72}\\
\texttt{ctrl} & 1.6259 & 1.3909 & \textbf{14.46} & \texttt{priority} & 14.3388 & 13.8604 & \textbf{3.34}\\
\texttt{div} & 570.7600 & 536.2190 & \textbf{6.05} & \texttt{router} & 4.6807 & 3.9507 & \textbf{15.60}\\
\texttt{i2c} & 3.6013 & 3.4579 & \textbf{3.98} & \texttt{sin} & 25.7537 & 24.5111 & \textbf{4.82}\\
\texttt{log2} & 53.0293 & 51.4379 & \textbf{3.00} & \texttt{sqrt} & 588.2180 & 542.3540 & \textbf{7.80}\\
\texttt{max} & 23.2156 & 21.0590 & \textbf{9.29} & \texttt{square} & 25.8996 & 24.1181 & \textbf{6.88}\\
\texttt{mem\_ctrl} & 18.5036 & 18.1849 & \textbf{1.72} & \texttt{voter} & 14.9015 & 13.6263 & \textbf{8.56}\\
\midrule
\multicolumn{7}{l}{Geometric-mean CPD reduction (\%)} & 6.82\\
\bottomrule
\end{tabular}
\end{table*}

\subsection{Combining greedy refinement and cut proposals}
Greedy refinement explores local combinations of sink assignments and cuts, while minimum-cut proposals coordinate replication across all current critical paths. Algorithm~\ref{alg:fapo} combines these complementary searches.

Algorithm~\ref{alg:fapo} maintains a \emph{search frontier} $\mathcal F$ of mappings to expand and a \emph{result archive} $\mathcal A$ of retained candidates. Greedy expansion advances the frontier. The archive stores a bounded set of nondominated area--delay tradeoffs. When $\alpha>0$, $\beta>0$, and $\mathcal Q$ is nonempty, $\operatorname{CutBundle}$ starts from the minimum-ATP mapping in a bounded Pareto sample. A legal, area-feasible result enters the archive only if it strictly reduces delay relative to that base. Each sweep contributes at most one cut proposal, which may contain multiple copy actions. Only the greedy branch advances the frontier.

At termination, final area recovery precedes minimum-ATP selection from the retained mappings and the protected seed.

\section{Experimental Evaluation}
\subsection{Experimental Setup}
The proposed FAPO framework is implemented in C++. All experiments ran on an Intel Xeon Gold 6230 CPU operating at 2.4~GHz with 256~GB of memory.

We used the EPFL benchmark suite \cite{AmaruEtAl2015EPFL}. ABC \texttt{if}, ACD-remap, and AGDMap-D provided the baseline mappings \cite{BraytonMishchenko2010ABC,TempiaCalvinoEtAl2025ACD,FanWu2023AGDMap}. Placement and routing were performed using VPR from VTR~9.0 with a 40-nm FPGA architecture \cite{ElgammalEtAl2025VTR9}.

\subsection{Comparison with Initial Mappings}
Unless otherwise stated, mapping experiments used $K=6$, $\alpha=1$, and $\beta=5\%$. ABC \texttt{if} used the command sequence \texttt{strash; balance; if -K 6}. To assess applicability across mapping flows, we applied the same FAPO policy to all three baseline mappers. We selected minimum-ATP mappings within each circuit's final-area cap. Table~\ref{tab:parents} reports per-circuit results and geometric-mean ATP ratios. Geometric-mean ATP decreased by 12.02\%, 14.48\%, and 12.04\% for ABC \texttt{if}, ACD-remap, and AGDMap-D, respectively. LUT-count increases remained within the 5\% allowance for every circuit.

\subsection{Search-Strategy Ablation}
To assess minimum-cut proposals, we compared five strategies on eight ACD-remap development circuits (Table~\ref{tab:ablation}). Greedy-only achieved an ATP ratio of 0.8607, compared with 0.8757 for minimum-cut-only. Combining the branches in FAPO reduced the ratio to 0.8553. We then evaluated FAPO on the remaining circuits. Relative to matched greedy-only search, FAPO achieved ATP ratios of 0.9938 on the development set and 0.9941 on the held-out set, respectively, yielding an overall ATP reduction of 0.60\%.

\subsection{Effect of the Final-Area Allowance}
\begin{figure}[t]
\centering
\includegraphics[width=\columnwidth]{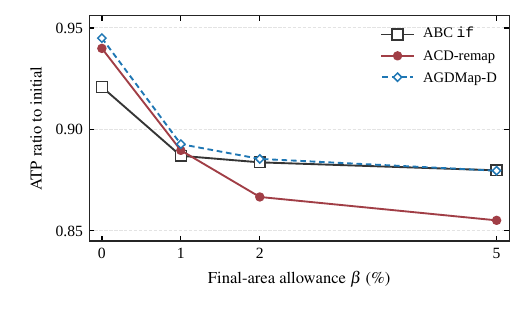}
\caption{Area-allowance sensitivity of FAPO. Each curve reports geometric-mean ATP ratios to its baseline mapper on EPFL benchmarks; lower is better.}
\label{fig:budget}
\end{figure}

To measure sensitivity to available area, we swept $\beta\in\{0,1,2,5\}\%$ for all three baseline mappers, selecting minimum ATP independently at each allowance. Without net area growth, ATP decreased by 5.50\%--7.91\% relative to the initial mappings (Fig.~\ref{fig:budget}). A 1\% allowance increased these reductions to 10.73\%--11.30\%. For ACD-remap, reductions reached 13.33\% at 2\% allowance and 14.48\% at 5\%, showing diminishing gains from additional area.

\subsection{Post-Route Evaluation}
To validate the physical timing benefits of FAPO, we evaluated the circuits with strict mapping-level ATP improvements over ABC \texttt{if}. The runs used full FAPO with $\beta=5\%$ on an FPGA architecture with 6-input LUTs and 10 fracturable logic elements per configurable logic block (CLB). Both variants used identical devices, channel widths, and placement-and-routing settings. Table~\ref{tab:physical} reports a 6.82\% geometric-mean reduction in \emph{critical-path delay} (CPD). At the selected seeds, geometric-mean refined-to-initial ratios were 1.0012 for packed CLB count and 1.0097 for wirelength.

\section{Conclusion}
FAPO coordinates sink assignments and LUT cuts under a final live-LUT area constraint. Full-network evaluation accounts for the coupled effects of replication and recovery, while minimum-cut proposals complement greedy refinement across tied critical paths. On EPFL benchmarks with three baseline mappers, FAPO reduces geometric-mean ATP by up to 14.48\% within a 5\% area allowance. VPR results show a 6.82\% routed-delay reduction on the circuits with mapping-level improvements.
\section*{Acknowledgment}
The authors used OpenAI Codex to assist with language editing in Sections I--IV. All changes were reviewed and verified by the authors.

\clearpage
\IEEEtriggeratref{11}

\end{document}